\documentclass[aps,pra,reprint,floatfix]{revtex4-1}

\usepackage{amsmath,amsfonts,amssymb}
\usepackage{graphicx}
\usepackage{color}
\usepackage{ulem}
\usepackage{float}
\usepackage{booktabs}

\def\beq{\begin{equation}}
\def\eeq{\end{equation}}
\def\beqn{\begin{eqnarray}}
\def\eeqn{\end{eqnarray}}

\begin{document}

\title{Efficient non-resonant intermolecular vibrational energy transfer}

\author{Lorenz S. Cederbaum}
\affiliation{Theoretische Chemie, Physikalisch--Chemisches Institut, Heidelberg University, 
Im Neuenheimer Feld 229, D-69120 Heidelberg, Germany}
\date{\today}

\begin{abstract}
Molecular excited vibrational states are metastable states and we incorporate their finite lifetimes into the theory of vibrational energy transfer between weakly interacting molecules, i.e., at internuclear distances at which they do not have a chemical bond. Expressions for the effective lifetime of an initially vibrationally excited molecule in the presence of a neighboring molecule are derived in closed form. These expressions allow one to analyze the physics behind the energy transfer. It is shown that due to different finite lifetimes of the isolated excited molecules, a very efficient vibrational energy transfer can take place between them even if their energies are rather off-resonance. Examples are discussed.  
\end{abstract}

\maketitle

Vibrations constitute a fundamental property of molecules and of other matter made of molecules. Excited vibrational molecular levels in the electronic ground state decay rather slowly radiatively, their radiative lifetime is typically in the range of seconds to milliseconds \cite{vib_Radzig_Smirnov}, see also \cite{rad_lifetime_Tennyson}. In contrast, fast non-radiative lifetimes in the range of pico- and even femtoseconds have been reported for even small polyatomic molecules \cite{vib_Quack,vib_Lehmann,vib_Albert}. The underlying mechanism of intramolecular vibrational energy redistribution (IVR) has attracted much attention for several decades \cite{vib_Quack,vib_Lehmann,vib_Albert,vib_Uzer,vib_Nesbitt,vib_Abel_Troe}. Here, an optically active vibrational mode couples to other modes which in turn couple to other modes (tier model \cite{vib_Sibert,vib_Albert}) redistributing the energy of the initially excited mode over other parts of the molecule. Anharmonic coupling plays here a decisive role. We shall return to the IVR below. 

The presence of IVR suggests the investigation of the more general subject of vibrational energy flow in molecules which has become an active field of research (for a recent review see \cite{vib_David}). Multidimensional spectroscopy experiments and theoretical investigations provide valuable information on this flow and hence on vibrational energy transport (or transfer) in condensed phase molecular systems, see, e.g., \cite{vib_Leitner_Straub,vib_Rubtsova,vib_Ishikura,vib_Uri,vib_Stock} and references therein. The major driving force of vibrational energy transfer is again anharmonicity \cite{vib_Uri,vib_Stock,vib_David2}. 

In this work we investigate the possibility for vibrational energy transfer between weakly interacting molecules, i.e., between molecules at distances where chemical bonds are not formed. We show that indeed such a transfer can be efficient even if it is not resonant, i.e., the respective levels are rather far from matching. The main ingredient is that the participating vibrational excited levels are metastable states of finite lifetimes. These lifetimes can be due to radiative decay, IVR or any other kind of decay, like predissociation. Consider a molecule 1 with a vibrational state of complex energy $E_1 - i\Gamma_1/2$ which interacts with a vibrational state of a neighboring molecule 2 of complex energy $E_2 - i\Gamma_2/2$, where $\Gamma$ is the width of a level of lifetime $\tau = \hbar/\Gamma$ \cite{res_Nimrod,res_Robin}. The time-evolution of the coupled metastable states (also called resonances) is governed by the Schroedinger equation \cite{res_Nimrod,res_Robin,DICES}
\begin{align}
\label{eq::1}
i\hbar\left(
\begin{array}{c}
\dot{\phi_1} \\
\dot{\phi_2} \\
\end{array}
\right)
=  \left(
\begin{array}{cc}
{E_1 - i\Gamma_1/2} & {W} \\
{W} & {E_2 - i\Gamma_2/2} \\
\end{array} 
\right)
\left(
\begin{array}{c}
\phi_1 \\
\phi_2 \\
\end{array}
\right),
\end{align}
where, as usual, the dot implies derivation with respect to time and $W$ is the coupling strength discussed below.

As we shall see below, in spite of its simple appearance, the physical content of the above equation is very rich. Therefore, we would like to approach the full complexity of the solution step by step. Let us begin by populating at time $t=0$ the excited vibrational level of molecule 1 while molecule 2 is not excited. If the two molecules are identical, the outcome is rather trivial: ${\phi}^{2}_1(t) = \exp(-t/\tau){\cos}^{2}(Wt/\hbar)$. Clearly, the population oscillates between the molecules with a frequency given by $W/\hbar$. In the general case the time evolution is rather involved and we would like to attribute a single number to the problem characterizing the {\it effective lifetime $\tau_{eff}$} of molecule 1 due to the presence of molecule 2. This is defined as the equivalent of the lifetime of an exponential decaying population $P_1 = exp(-t/\tau_{eff})$ :
\begin{align*}
\nonumber
\tau_{eff} = \int\limits_{0}^{\infty} P_1 dt.
\end{align*}

Returning to the simple case of two identical molecules, one immediately finds
\begin{align}
\label{eq::2a}
\tag{2a}
\tau_{eff} = \frac{\hbar}{\Gamma}\,\frac{2W^{2} + \Gamma^{2}}{4W^{2} + \Gamma^{2}}.
\end{align}

If the coupling $W^{2}$ is very small compared to $\Gamma^{2}$, one retains the original lifetime $\tau$, and if this coupling is very large, one gets $\tau_{eff} = \tau/2$. Assuming that the decay is radiative and the experiment repeated many times, the total number of photons emitted from both molecules is, of course, the same as emitted from molecule 1 alone in the absence of the neighbor. But if one can measure the photons emitted from molecule 1 in the presence of the neighbor ($W^{2}$ large), one finds only half the photons. 

Next, we consider two molecules which are not in resonance, $\Delta = E_1 - E_2 \neq 0$, but still posses the same lifetime $\tau$. The calculation is again trivial because the eigenvectors of the Hamiltonian matrix in Eq.(\ref{eq::1}) are real. The result reads:
\begin{align}
\label{eq::2b}
\tag{2b}
\tau_{eff} = \frac{\hbar}{\Gamma}\,\frac{\Delta^{2} + 2W^{2} + \Gamma^{2}}{\Delta^{2} + 4W^{2} + \Gamma^{2}}.
\end{align}
\stepcounter{equation}

Now, one has to compare $W^{2}$ with $\Delta^{2} + \Gamma^{2}$, and as we shall see below both the coupling and width are rather small in reality, the effective lifetime approaches that of the isolated molecule very rapidly with growing energy mismatch $\Delta$. Consider, e.g., $W=\Gamma=0.1 cm^{-1}$ which are generous numbers for radiative decay (see below for details), then for two identical molecules we have $\tau_{eff} = 0.6 \tau$ which is a considerable reduction of the lifetime of molecule 1 due to energy transfer to its neighbor. This reduction, however, essentially disappears for $\Delta$ as small as $1cm^{-1}$. A Taylor expansion of Eq.(\ref{eq::2b}) readily gives $(\tau-\tau_{eff})/\tau = 2W^{2}/\Delta^{2}$ as the leading term. This is, in general, the reason why in energy transfer situations one is considering resonant energy transfer, see, e.g., \cite{elec_Scholes}. 

We now come to our main objective and show that the situation can be dramatically different if one allows for the lifetimes of the two molecules to be different, i.e., $\Gamma_{1} \neq \Gamma_{2}$.

The Schroedinger equation (\ref{eq::1}) can be solved explicitly because the matrix Hamiltonian does not depend on time. Diagonalizing this matrix provides complex eigenvalues 
\begin{align}
\label{eq::3}
\lambda_\pm = E_\pm - i\Gamma_{\pm}/2,
\end{align}
and complex eigenvectors with elements $U_\pm, V_\pm$. The eigenvector matrix ${\bf U}$ fulfills the usual normalization condition for a complex symmetric matrix: ${\bf U}^{T}{\bf U} = {\bf 1}$, where $T$ stands for transposed. The resonance state $\phi_{1}(t)$ with the initial condition $\phi_{1}(0)=1$ follows readily when using the normalization $U^{2}_\pm + V^{2}_\pm = 1$:
\begin{align}
\label{eq::4}
\phi_{1} = U^{2}_{+} e^{-iE_{+}t/\hbar}e^{-\Gamma_{+}t/2\hbar} + U^{2}_{-} e^{-iE_{-}t/\hbar}e^{-\Gamma_{-}t/2\hbar}.
\end{align}
The population $P_{1}$ of the resonance as a function of time is computed as the $f$-product of the resonance state  \cite{f_product_Nimrod} and integrating it over time provides after some minor manipulations the effective lifetime of the resonance:
\begin{align}
\label{eq::5}
\tau_{eff} = U^{4}_{+} \frac{\hbar}{\Gamma_{+}} + U^{4}_{-} \frac{\hbar}{\Gamma_{-}} + 2U^{2}_{+}U^{2}_{-}\frac{\hbar\bar{\Gamma}}{(E_{+}-E_{-})^{2} + \bar{\Gamma}^{2}},
\end{align}
where $\bar{\Gamma} = (\Gamma_{1} + \Gamma_{2})/2$ is the average width of the two molecules. Note that because of the normalization $U^{2}_{+} + U^{2}_{-} = 1$, one can view the above expression as a weighted sum of various lifetimes. 

The explicit calculation of the eigenvectors and eigenvalues is straightforward, but rather lengthy. For computing $E_\pm$ and $\Gamma_\pm$ one needs to solve for the real and imaginary parts of $\sqrt{\Delta^{2} - (\delta/2)^{2} + 4W^{2} + i\Delta^{2}\delta^{2}}$, where we have introduced $\delta = \Gamma_{2} - \Gamma_{1}$. For $\Delta\delta>0$ we obtain
\begin{align*}
\nonumber
&E_{+} - E_{-} = \sqrt{A + \sqrt{A^{2} + \Delta^{2}\delta/2)^{2}}}/ \sqrt{2},\\
&\Gamma_{+} - \Gamma_{-} = -\sqrt{-A + \sqrt{A^{2} + \Delta^{2}\delta/2)^{2}}}\sqrt{2},\\
&A = \Delta^{2} - (\delta/2)^{2} + 4W^{2},
\end{align*}
and the same result except of a sign is found for $\Delta\delta<0$. Noting that $E_{+} + E_{-} = E_{1} + E_{2}$ and $\Gamma_{+} + \Gamma_{-} = \Gamma_{1} + \Gamma_{2}$, one readily finds the individual energies and widths of the resulting resonances. The eigenvector components follow from the above explicit expressions as usual. The expressions for the eigenvectors are lengthy. Interestingly, when inserted into Eq.(\ref{eq::5}), many terms cancel or add up nicely to give an explicit formula for $\tau_{eff}$ which will be discussed in the following. 

We start the discussion with the case where the vibrational levels of the two molecules are degenerate, but their lifetimes are different. One can learn much from this case which is an example contrasting the trivial case addressed in Eq.(\ref{eq::2a}) where the energies as well as the lifetimes of the two molecules are the same. Since $\Delta = E_{1} - E_{2} = 0$ in our case, we see from the expression above that for $W^{2} = (\Gamma_{1} - \Gamma_{2})^{2}/16$ the two complex energies become degenerate, i.e., the resonances coalesce. This is known to lead to a branch point in the complex energy plane and to self-orthogonality of the resonance which in turn gives rise to interesting physics \cite{res_Nimrod,res_Robin}. We will not dwell on this issue here because the case $\Delta = 0$ is not general enough. Here, we just mention that due to the branch point we obtain different expressions for the effective lifetime depending on whether $W^{2}$ is larger or smaller than $(\Gamma_{1} - \Gamma_{2})^{2}/16$. 

Evaluating Eq.(\ref{eq::5}) leads to the following explicit expressions
\begin{align}
\label{eq::6}
&\tau_{eff} = \frac{\hbar}{\bar{\Gamma}}\,\frac{2W^{2} + \bar{\Gamma}\Gamma_{2}}{4W^{2} + \Gamma_{1}\Gamma_{2}} \qquad for \quad W^{2} < \frac{(\Gamma_{1} - \Gamma_{2})^{2}}{16} \\
&\nonumber \tau_{eff} = \frac{\hbar}{\bar{\Gamma}}\,\frac{2W^{2} + \Gamma_{1}\Gamma_{2}}{4W^{2} + \Gamma_{1}\Gamma_{2}} \qquad for \quad W^{2} > \frac{(\Gamma_{1} - \Gamma_{2})^{2}}{16}.
\end{align}
Having explicit compact expressions at hand, one can see the richness of the underlying physics. Note that average width $\bar{\Gamma}$ appears in the prefactor in the above expressions. In contrast to the trivial case addressed in Eq.(\ref{eq::2a}) where the effective lifetime can only vary within a factor of $2$, the range of variation in Eq.(\ref{eq::6}) is very large and can cover several orders of magnitude. If the coupling $W^{2}$ between the molecules is large compared to $\Gamma_{1}\Gamma_{2}$, we see that the resonance acquires the total width $\Gamma_{1} + \Gamma_{2}$. This is particularly interesting in the case where the initially excited molecule 1 has a very long lifetime $\tau_{1}$ and the neighboring molecule a short lifetime $\tau_{2}$. Then, $\Gamma_{1}\Gamma_{2}$ is indeed small and the energy transfer is very efficient as indicated by the smallness of $\tau_{eff}$. Although molecule 1 lives long when isolated, it decays very fast in the neighborhood of molecule 2. We shall return to this scenario later. 

What happens if the vibrational levels of the two molecules are off resonance, i.e., $\Delta = E_1 - E_2 \neq 0$? In the trivial case of equal lifetimes discussed around Eq.(\ref{eq::2b}), the efficiency of the energy transfer decreases fast with increasing $\mid\Delta\mid$. We shall see that this is not at all the case when the lifetimes are different. The general expression for the effective lifetime takes on the following appearance:
\begin{align}
\label{eq::7}
&\tau_{eff} = \frac{\hbar}{\bar{\Gamma}}\,\frac{B + D/C}{4W^{2} + \Gamma_{1}\Gamma_{2}(\Delta^{2}+\bar{\Gamma}^{2})/\bar{\Gamma}^{2}}, \\
&\nonumber C =  \sqrt{A^{2} + \Delta^{2}\delta^{2}}, \\
&\nonumber D =  [2\Delta^{4} + \Delta^{2}(8W^{2} + \delta\Gamma_{2} + \bar{\Gamma}\delta) - \bar{\Gamma}\delta A]\Gamma_{2}/4\bar{\Gamma}, \\
&\nonumber B = \Delta^{2}\Gamma_{2}/2\bar{\Gamma} + 2W^{2} + \Gamma_{2}(\Gamma_{2} + 3\Gamma_{1})/4.
\end{align}
The quantities $A$ and $\delta$ have been defined above. Of course, this expression is not as compact as that in Eq.(\ref{eq::6}), but it is still of simple structure and easy to evaluate.

In contrast to the expression for $\tau_{eff}$ in Eq.(\ref{eq::2b}) where the energy gap enters the denominator additively as $\Delta^{2}$, it enters in Eq.(\ref{eq::7}) as $\Delta^{2}\Gamma_{1}\Gamma_{2}/\bar{\Gamma}^{2}$. In the scenario in which $\Gamma_{1}$ is very small compared to $\Gamma_{2}$ we thus have a wide window of the energy gap where energy transfer is efficient. To see this, we assume that the energy gap is much larger than the coupling, i.e., $\Delta^{2} \gg W^{2}$, and at the same time $W^{2} \gg \Delta^{2}\Gamma_{1}\Gamma_{2}/4\bar{\Gamma}^{2}$. A Taylor expansion of $\tau_{eff}$ in Eq.(\ref{eq::7}) under this condition gives: $\tau_{eff} \approx \frac{\hbar}{\Gamma_{2}}\,\frac{\Delta^{2}}{W^{2}}$. We now multiply in this expression $\Delta^{2}$ as well as $\Gamma_{2}$ by $\Gamma_{1}\Gamma_{2}/4\bar{\Gamma}^{2}$, and because $\Gamma_{2}^{2}/4\bar{\Gamma}^{2}$ is essentially equal to $1$, we find $\tau_{eff} \approx \frac{\hbar}{\Gamma_{1}}\,\frac{\Delta^{2}\Gamma_{1}\Gamma_{2}/4\bar{\Gamma}^{2}}{W^{2}} \ll \frac{\hbar}{\Gamma_{1}}$. I.e., the effective lifetime is much less than the lifetime of the isolated molecule 1 due to efficient vibrational energy transfer to molecule 2 in spite of a large energy gap.

For completeness, we very briefly discuss some limiting cases of Eq.(\ref{eq::7}). For very large energy gap $\mid\Delta\mid \rightarrow\infty$, we notice that $C \rightarrow\Delta^{2}$ and collecting all leading terms in the numerator and denominator in the equation, one immediately gets $\tau_{eff}\rightarrow \frac{\hbar}{\bar{\Gamma}}\, \frac{\Delta^{2}\Gamma_{2}/\bar{\Gamma}}{\Delta^{2}\Gamma_{1}\Gamma_{2}/\bar{\Gamma}^{2}} = \frac{\hbar}{\Gamma_{1}}$. As expected, the lifetime is that of the isolated molecule 1. For a vanishing gap $\Delta = 0$, one finds that $C = \mid- (\delta/2)^{2} + 4W^{2}\mid$, implying that $C$ takes on different expressions depending on whether $4W^{2}$ is larger or smaller than $(\delta/2)^{2}$. Inserting these expressions into Eq.(\ref{eq::7}), one arrives at the results (\ref{eq::6}). Finally, for $\mid W\mid \rightarrow\infty$, we collect all leading terms, $B$ gives $2W^{2}$, $D/C$ does not contribute and we simply obtain $\tau_{eff}\rightarrow \frac{\hbar}{\Gamma_{1} + \Gamma_{2}}$. I.e., in this limit the effective width is the sum of both widths. 

Intermolecular {\it electronic} energy transfer is often described by expanding the electron-electron Coulomb interaction in terms of the separation $R$ between the molecules, see, e.g., \cite{elec_Scholes,res_Robin}. The resulting coupling matrix elements between the involved electronic states is that of a transition dipole-transition dipole interaction and is much used in various contexts like, for instance, in exciton transfer in semiconductors \cite{exciton_Fink}. One can use the same scheme for intermolecular {\it vibrational} energy transfer by taking the matrix elements of the long-range electron-electron interaction between the involved vibrational states. This leads to the same expression for the coupling, but the dipole transitions are now those of the vibrational states. This idea is not new and has been used before for describing resonant transfer in the condensed phase, see, e.g., \cite{RVET_Water_Woutersen,VET_condensed_Chen}.

We shall use here the same coupling, but as discussed above in the context of non-resonant transfer between weakly interacting molecules at internuclear distances at which chemical bonds are not present. We express the coupling in quantities and units suitable for vibrational states:
\begin{align}
\label{eq::8}
W^{2}[cm^{-1}]^{2} =  2.58 \times 10^{20} \, \frac{A_{1}[s^{-1}]\,A_{2}[s^{-1}]} {\nu_{1}^{3}[cm^{-1}]^{3}\,\nu_{2}^{3}[cm^{-1}]^{3}} \,\frac{\alpha}{R^{6}[{{\AA}}]^{6}}.
\end{align}
Here, $A_{1}$ is the Einstein coefficient of the vibrational transition of frequency $\nu_{1}$ from the excited state to the ground (or another lower lying) state of molecule 1 and similarly for molecule 2. $\alpha$ is a number factor arising due to the relative orientation of the dipoles of the two molecules. $\alpha=4$ if the two dipoles are parallel or antiparallel and $\alpha=2/3$ if $W^{2}$ is averaged over a random orientation of the molecules. The Einstein coefficient $A_{1} = 1/\tau_{1}^{\it rad}$ is related to the radiative lifetime $\tau_{1}^{\it rad}$ of a transition. For completeness we also give the relation between width and lifetime in appropriate units:
\begin{align}
\label{eq::9}
\tau [s] = \frac{5.31 \times 10^{-12}}{\Gamma [cm^{-1}]}.
\end{align}

We now turn to examples. According to our analytic result (\ref{eq::7}), starting from a long-lived level of molecule 1, resonant and, in particular, non-resonant vibrational energy transfer is the more efficient the shorter the lifetime of the neighboring molecule 2 is. Obviously, a strong coupling $W$ helps. For a large $W$ one needs low vibrational frequencies and short radiative lifetimes. IVR and other mechanisms lead to very short (non-radiative) lifetimes, much shorter than radiative ones. Consequently, we choose as our first example a neighbor with efficient IVR. There are many such molecules \cite{vib_Quack,vib_Lehmann,vib_Albert,vib_Uzer,vib_Nesbitt,vib_Abel_Troe}, and we choose $^{13}CHF_{3}$ which been remeasured accurately recently and well understood theoretically \cite{IVR_CHF3_Albert} as a representative case. Due to a Fermi resonance, there is fast energy flow between the fundamental stretching (3024.6 $cm^{-1}$) and bending of the CH-chromophore (100 fs) and the exchange of energy with the heavy-atom frame has 11 ps as its time scale \cite{IVR_CHF3_Albert} which we choose as the lifetime of molecule 2. The radiative lifetime of the fundamental CH vibration can be deduced from \cite{IVR_CHF3_Marquardt} and is 18 ms. These are rather typical radiative lifetimes and frequencies, and we compute the coupling using them also for molecule 1, obtaining
\begin{align*}
\nonumber
W [cm^{-1}] = 64.5/R^{3}[{{\AA}}]^{3},
\end{align*}
which gives $W=1 cm^{-1}$ at a distance of $R=4{{\AA}}$, $0.52 cm^{-1}$ at $5{{\AA}}$ and $0.064 cm^{-1}$ at a distance of a full nanometer. We shall see below that such couplings lead to substantial effects. 

Note that in general one has an enormous choice for molecule 1. For instance, diatomic molecules of similar frequencies are good choices, as their lifetimes are purely radiative \cite{Frequencies_Diatomic_Huber_Herzberg}. In Fig.1 the effective lifetime of molecule 1 computed with Eq.(\ref{eq::7}) is shown for the above parameters as a function of the energy mismatch $\Delta$. The widths $\Gamma_{1}=2.95\times10^{-10} cm^{-1}$ and $\Gamma_{2}=0.483 cm^{-1}$ corresponding to the lifetimes $\tau_{1}=18 ms$ and $\tau_{2}=11 ps$, respectively, have been used. At resonance, i.e., $\Delta=0$, one immediately obtains via Eq.(\ref{eq::6}) a reduction of the lifetime of molecule 1 due to the presence of the neighbor from $18ms$ to $11ps$ for $W = 1 cm^{-1}$ and $0.52 cm^{-1}$, and to $167ps$ for $W = 0.064 cm^{-1}$. From the figure we see that the lifetimes become longer as $\mid\Delta\mid$ increases, but the energy transfer remains efficient: At $\mid\Delta\mid$ as large as $50 cm^{-1}$ the lifetime of molecule 1 still reduces dramatically to around 30ns, 150ns and to less than $10\mu s$ for the above values of $W$. Even at an energy split as large as $100 cm^{-1}$ the reduction is still substantial. It is surprising to see that at distance of $1nm$ and an energy mismatch of $100 cm^{-1}$, the lifetime is still less than $30\mu s$.  

\begin{figure}[!]
	\includegraphics[width=1.0\columnwidth,angle=0]{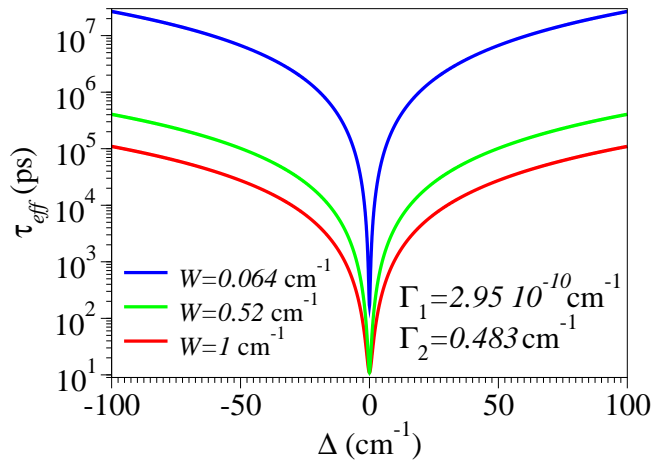}
	\caption{(Color online) The effective lifetime $\tau_{eff}$ of molecule 1 in the presence of a neighboring molecule 2 as a function of the energy difference $\Delta$ between the vibrational levels of the two molecules. The three curves are for different values of the coupling $W$, corresponding to distances of $R= 4$, $5$ and $10 {{\AA}}$. The lifetime of the isolated molecule 1 is $18 ms$. It is seen that this lifetime can reduce dramatically due to the neighbor even for large energy mismatch. See main text for details.}
	\label{fig1}
\end{figure}

Another 
class of systems where vibrational energy transfer is of interest are clusters of weakly bonded molecules. Experimental vibrational spectra of cluster are difficult to interpret without accurate theoretical analysis. Spectra have been recorded, e.g., for $(HF)_{n}$ clusters and analyzed by {\it ab initio} methods \cite{HF_n_Quack}. Infrared spectra of various mass-selected charged clusters were observed by vibrational predissociation spectroscopy, see, e.g., \cite{IR_Zundel_Okumura,IR_Zundel_Fournier}. For a quantum mechanical treatment of an interesting case, see \cite{IR_Zundel_Meyer}. Cluster dissociation by the loss of a weakly bonded tag-molecule or atom follows a vibrational excitation and the spectra are recorded by detecting fragment ions as a function of laser frequency. 
Recently, experimental spectra of $D_{2}$-tagged $H^{+}(H_{2}O)_{4}$ have been reported and discussed \cite{IR_Zundel_Exp_Johnson,IR_Zundel_Theory_Bowman}. The spectra exhibit broad peaks beyond the experimental resolution indicating 
very short sub-ps lifetimes probably due to predissociation or other kinds of vibrational energy redistribution. This would make such clusters suitable partners for efficient interspecies vibrational energy transfer from long-lived molecules in their vicinity or attached to them being another constituent of the cluster.  

Understanding the vibrations of adsorbates at surfaces and their relaxation is important for many applications in chemistry and physics \cite{D_on_Surface.Saalfrank_CP,Review_Surface.Arnolds}. Molecules adsorbed on metal surfaces may couple to electron-hole pairs of the surface and their non-radiative vibrational lifetimes are often in the ps range or even shorter due to vibration-electron coupling \cite{vet_electr_Tully,Review_Surface.Arnolds,vet_electr_CO_Tully}. 
 The lifetime of vibrations of adsorbates can be rather short (ns to sub-ps) also due to vibration-phonon coupling, in particular, when the vibration lies within the phonon band of the surface \cite{D_on_Surface.Saalfrank_CP,D_on_Surface.Saalfrank_JCP}. 

Following the mechanism discussed in this work, long-lived molecules in the gas phase in the vicinity of the surface can efficiently transfer their vibrational energy to adsorbates with similar vibrational frequencies. To be specific we consider D adsorbed on Si(100). The D$-$Si bending mode of frequency
$458 cm^{-1}$ has a lifetime of around $210 fs$ \cite{D_on_Surface.Saalfrank_JCP}. Using the data of \cite{H_on_Surface.Saalfrank_PRB}, its radiative lifetime can be estimated to be $0.99s$. Choosing a standard molecule with a frequency of $500 cm^{-1}$ and a lifetime of $1s$ in the vicinity of the surface, the coupling $W [cm^{-1}] = 296/R^{3}[{{\AA}}]^{3}$ results. At $R = 1 nm$ the lifetime of this molecule reduces due to the energy transfer from $1s$ to $\tau_{eff} = 4.6 ns$ in spite of the detuning $\Delta = 42 cm^{-1}$. In another scenario, the long-lived molecule could also be an adsorbed molecule which is physisorbed and not chemisorbed to the surface of the solid and hence will have much weaker vibration-phonon coupling and thus maintain its long-lifetime.   

In conclusion, there is efficient vibrational energy transfer between a vibrationally excited long-lived molecule and a neighbor which would be short-lived when excited. The two molecules can be rather off-resonance and still there is a substantial transfer effect. Because of the generality of the effect, many applications are possible. Two remarks are in order. If the long-lived molecule has several short-lived neighbors, the effective lifetime will reduce further. For $N$ equivalent neighbors, $\tau_{eff}\rightarrow  \tau_{eff}/N$, see \cite{res_Robin}. The neighbor does not have to initially be in its ground state, it could also be in an excited long-lived state and be transferred by the effect discussed here to a higher excited short-lived level.


\section*{Acknowledgements} 
The author thanks I. Baldea, S. Klaiman, A. Kuleff and R. Marquardt for valuable contributions. Financial support by the DFG (research unit 1789) and by the European Research Council (ERC) (Advanced Investigator Grant No. 692657) is gratefully acknowledged



\end{document}